\documentclass{iopart}
\usepackage{iopams}
\usepackage{graphicx,graphics}
\usepackage{epsf}
\usepackage{epsfig}
\usepackage{dcolumn}
\usepackage{bm}
\usepackage{rotate}
\usepackage[squaren,thickqspace,mediumspace]{SIunits}
\usepackage{subfigure}
\usepackage{longtable}
\usepackage{lscape}
\usepackage{multirow}
\usepackage{relsize}

\begin{document}
\title{Universal Window for Two Dimensional Critical Exponents}

\author{A Taroni$^1$\footnote{Present address: Department of Physics, Uppsala University, Box 530, 751 21 Uppsala, Sweden}, S T Bramwell$^{2,1}$ and P C W Holdsworth$^3$}

\address{$^1$ University College London, Department of Chemistry, 20 Gordon Street, London WC1H~0AJ, UK}
\address{$^2$ London Centre for Nanotechnology, 17-19 Gordon Street, London WC1H~0AH, UK}
\address{$^3$ Universit\'{e} de Lyon, Laboratoire de Physique, \'{E}cole Normale Sup\'{e}rieure de Lyon, 46 All\'{e}e d'Italie, 69364 Lyon cedex 07, France}
\eads{andrea.taroni@fysik.uu.se, s.t.bramwell@ucl.ac.uk, peter.holdsworth@ens-lyon.fr}

\begin{abstract}
Two dimensional condensed matter is realised in increasingly diverse forms that are accessible to experiment and of potential technological value. The properties of these systems are influenced by many length scales and reflect both generic physics and chemical detail. To unify their physical description is therefore a complex and important challenge. Here we investigate the distribution of experimentally estimated critical exponents, $\beta$, that characterize the evolution of the order parameter through the ordering transition. The distribution is found to be bimodal and bounded within a window $\sim 0.1 \le \beta \le 0.25$, facts that are only in partial agreement with the established theory of critical phenomena. In particular, the bounded nature of the distribution is impossible to reconcile with existing theory for one of the major universality  classes of two dimensional behaviour - the XY model with four fold crystal field - which predicts a spectrum of non-universal exponents bounded only from below. Through a combination of numerical and renormalization group arguments we resolve the contradiction between theory and experiment and demonstrate how the ``universal window'' for critical exponents observed in experiment arises from a competition between marginal operators.
\end{abstract}

\submitto{\JPCM}
\maketitle

\newpage
\section{Introduction}\label{survey}

New types of two-dimensional system on which meaningful physical experiments can be performed  include optical lattices of trapped atomic gases~\cite{Hadzibabic06}, magnetic surfaces~\cite{Rose07}
and ``$\delta$-doped" magnetic layers~\cite{Parnaste07}. These add to a list of well established two dimensional systems that includes ultrathin magnetic films~\cite{Elmers95}, atomic monolayers (both physi- and chemisorbed)~\cite{Wang85,Capunzano85,Bishop78,Nuttall95}, crystalline surfaces~\cite{Yang90}, superconducting layers~\cite{Choy98} and arrays of interacting Josephson junctions~\cite{Resnik81}.  Recent theoretical developments on the concept of ``extended universality''~\cite{Wexler06}, the effects of finite size~\cite{Chung06,Trombettoni05}, and the dipolar interaction~\cite{Maier04,Debell00} should be particularly relevant to understanding experiments on these systems, both old and new.

The key experiment on two dimensional systems is to test the existence and temperature dependence of a magnetic or crystalline order parameter $m(T)$. In cases where $m$ can be measured experimentally (which excludes, for example, superfluid films~\cite{Bishop78}), this is invariably found to approximate a power law over a certain range of temperature: $m \sim (T_{\mathrm{c}} - T)^{\beta}$, where $T_{\mathrm{c}}$ is the transition temperature. Theory predicts a limited number of possibilities for the value of the exponent $\beta$, as dictated by the universality class of the system. In two dimensions crystal symmetries and consequent universality classes are relatively few. We show here that the Ising, XY and XY with 4-fold crystal field anisotropy (XY$h_4$) are the three main experimentally relevant classes. The three and four state Potts models provide additional universality classes observed in experiments on adsorbed gaseous monolayers~\cite{Pfnur89,Pfnur90} and surface reconstruction~\cite{Yang90}. For the Ising, three- and four-state Potts models, $\beta = \frac{1}{8}, \frac{1}{9}, \frac{1}{12}$ respectively. For the XY model, one expects $\tilde{\beta} = 0.23$, a universal number that arises in the finite size scaling at the Kosterlitz-Thouless-Berezinskii (KTB) phase transition~\cite{Berezinskii70,Kosterlitz73}, though not a conventional critical exponent~\cite{Bramwell93}. For XY$h_4$, theory predicts a continuously variable critical exponent $\beta \propto 1/h_4$ and thus a continuous spectrum of values when sampled over many real systems (see references~\cite{Jose77,Calabrese02} and this work).

We have tested these ideas by means of an extensive survey of experimental two dimensional critical exponents, including data for magnetic ultrathin films, layered magnets that exhibit a temperature regime of two dimensional behaviour~\cite{Ikeda74,Als-Nielsen93,Hirakawa82,Regnault90}, order-disorder transitions in adsorbed gaseous monolayers~\cite{Pfnur89,Pfnur90}, and surface reconstructions~\cite{Yang90}. The results are presented in Figure \ref{histogram} and in the appendix. As observed previously on more limited data sets~\cite{Elmers95,Bramwell93,Bramwell93b,Qiu94}, the distribution of $\beta$'s is distinctly bimodal, with strong peaks at $\beta = 0.12$ and $\beta = 0.23$, as expected for the Ising and XY models. In several cases ideal Ising~\cite{Ikeda74,Elmers96_Ising} and XY~\cite{Als-Nielsen93,Hirakawa82,Regnault90,Thurlings89,Elmers96_XY} behaviour has been confirmed in great detail by measuring thermodynamic quantities other than the magnetization. Likewise there is compelling evidence for Potts universality in several non magnetic systems~\cite{Pfnur89,Pfnur90,Sokolowski94,Floreano01,Yang90}. However the XY$h_4$ universality is more elusive. Of particular relevance to the present discussion are the ferromagnetic monolayer Fe/W(100)~\cite{Elmers96_XY} and the layered ferromagnets Rb$_2$CrCl$_4$~\cite{Als-Nielsen93} and K$_2$CuF$_4$~\cite{Hirakawa82}, easy plane systems which have been shown to exhibit the full range of ideal XY behaviour despite their 4-fold symmetry. Another very well characterised easy plane system with a 4-fold crystal field is the layered antiferromagnet K$_2$FeF$_4$~\cite{Thurlings89}, but this is not XY-like, with $\beta = 0.15$ intermediate between the XY and Ising values. Claims for XY$h_4$ universality have been made for the ferromagnetic films Fe/[Au or Pd](100)~\cite{Rau96}, characterised only to a limited extent, as well as the order-disorder transitions of H/W(011)~\cite{Lyuksyutov81} and O/Mo(110)~\cite{Grzelakowski90}, for which full sets of critical exponents are available. The behaviour of these candidates for XY$h_4$ is seen to fall into two categories, which on closer inspection appears to be related to the strength of $h_4$: those with weak $h_4$ are XY-like with $\beta \approx 0.23$, while those with stronger $h_4$ have exponents in between the XY and Ising limits, $0.125 \le \beta \le 0.23$. Most strikingly, there is no experimental evidence of the divergence of the exponent $\beta$ implied by $\beta \propto 1/h_4$. Instead, most experimental data that cannot be ascribed to the Potts classes lies in a ``universal window'', bounded by the Ising and XY values. There are exceptions at the upper bound where crossover to three dimensional behaviour may increase the value of $\beta$ upwards from $0.23$~\cite{Bramwell93,Lancaster07}. However, it is clear from the histogram that the majority of systems are indeed encompassed in a limited range between the Ising and XY values.

\begin{figure}%[htp]
  \begin{center}
  \includegraphics[scale=0.45]{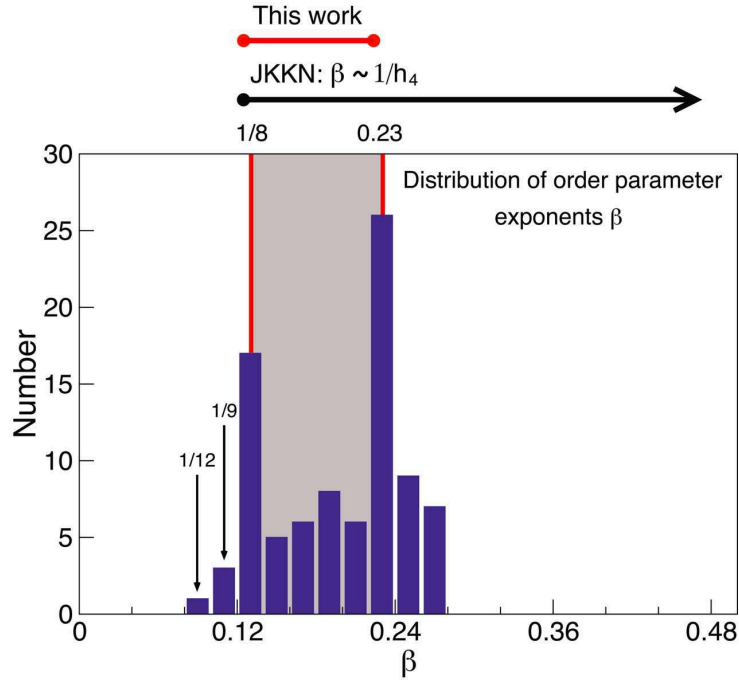}
  \caption{Histogram of $\beta$ values for all two-dimensional systems reported in Tables \ref{layered_table}, \ref{thinfilm_table} and \ref{orderdisorder}. The universal window is highlighted by the grey shading. Criteria for inclusion in the data set are discussed in the appendix.}
  \label{histogram}
  \end{center}
\end{figure}

\section{Universality Classes in Two Dimensions}

Before we address the main question of why the universal window exists, it is relevant to specify the occurrence and relationships between two dimensional universality classes. Considering first
magnetic degrees of freedom, we ignore the possibility of truly Heisenberg behaviour, remarking that the broken translational symmetry inherent to layers or surfaces, combined with a condition of crystal periodicity, means all real systems have at least one $p$-fold axis, which necessarily introduces relevant perturbations. Thus, although pure Heisenberg behaviour may be observable over a restricted temperature range~\cite{Ronnow98,Greven95}, it must give way to behaviour characteristic of the perturbations at temperatures near to the phase transition. These perturbations take the form of axial anisotropy (either easy axis or easy plane) and $p$-fold in-plane anisotropy ($p$ = 1,2,3,4 and 6). Easy-axis systems are generally Ising-like (despite the fact that the normal to the plane is usually a polar axis) while easy-plane systems with $p$ = 2-6 should be described by the XY$h_p$ model. XY$h_2$ is in the Ising class, whereas XY$h_3$ is in the 3-state Potts class, although it is very unlikely in magnetic systems owing to time reversal symmetry (we found no examples). XY$h_4$ constitutes a universality class distinct from the 4-state Potts class, while the phase transition in XY$h_6$ is in the XY class~\cite{Jose77}. Inclusion of the dipolar interaction on lattices other than the square lattice does not add extra universality classes. However the case of the square lattice must be regarded as an unsolved problem: perturbative calculations~\cite{Prakash90} and numerical results~\cite{Debell97,Carbognani02,Fernandez07} suggest that the square lattice dipolar model belongs to XY$h_4$, but the renormalisation group calculations of Maier and Schwabl indicate a different set of critical exponents~\cite{Maier04}. The experimental data considered here are consistent with the former result rather than with the latter, but Maier and Schwabl's prediction could yet be born out on an as yet undiscovered ideal model dipolar system. At least as far as the existing experimental data set is concerned, we conclude that, for magnetic systems, there are only three main universality classes: Ising, XY and XY$h_4$.

The situation is essentially the same in non-magnetic systems~\cite{Nelson79,Schick81} but with the additional possibility of the 3- or 4-state Potts classes due to competing interactions beyond nearest neighbour~\cite{Schick81,Landau85}. Indeed, Schick~\cite{Schick81} used arguments from Landau theory to classify the phase transitions of two dimensional adsorbed systems into only four classes: the Ising, XY$h_4$, 3- and 4-state Potts. This set is supplemented by a chiral 3-state Potts class which shares conventional exponents with the pure 3-state Potts class~\cite{Huse82} (hence for present purposes we shall treat these two cases as a single class). One result of the current work is that the pure XY class is also relevant to order-disorder transitions in adsorbed layers. Combining these observations we have five universality classes for structural systems and three for magnetic systems, as summarised in Table \ref{universality_classes}.

\begin{table}%[ht]
\caption{Classification of continuous transitions which can be observed in two-dimensional magnetic systems and in structural order-disorder transitions on surfaces. $\bullet$ indicates the occurrence of a particular universality class, whereas $\times$ indicates its absence. The special case of the square lattice dipolar system is discussed in the text.}
\vspace{1mm}
\begin{indented}
\item[]\begin{tabular}{ccc}
\hline
\hline
Universality Class & Magnetic Systems & Adsorbed Systems \\
\hline
Ising   & $\bullet$ & $\bullet$ \\
XY      & $\bullet$ & $\bullet$ \\
XY$h_4$ & $\bullet$ & $\bullet$ \\
3-state Potts & $\times$ & $\bullet$ \\
4-state Potts & $\times$ & $\bullet$ \\
\hline
\hline
\end{tabular}
\end{indented}
\label{universality_classes}
\end{table}

\section{Calculation of Critical Exponents}

The relationship between the Ising, XY, XY$h_p$ and clock models may be discussed with reference to the following Hamiltonian:
\begin{equation}\label{2DXY}
  \mathcal{H}_p= - J \sum_{\langle i,j\rangle} \cos(\theta_i - \theta_j) - h_p\sum_i \cos(p\theta_i),
\end{equation}
\noindent in which the $\theta_i$'s are the orientations of classical spins of unit length situated on a square lattice with periodic boundary conditions and confined to the XY plane, $J$ is the coupling constant and $h_p$ is the $p$-fold crystal field. It should be noted that unlike real systems, the lattice symmetry in computer simulations does not constrain the spin symmetry, and consequently the adoption of a square lattice does not restrict the generality of our arguments. In the limit $h_p\rightarrow\infty$, the Hamiltonian (\ref{2DXY}) is called a clock model, since $\theta_i$ is restricted to discrete values evenly spaced around a circle: $2\pi(n/p),\; n=1,\dots ,p-1$. Jos\'e, Kadanoff, Kirkpatrick and Nelson (JKKN)~\cite{Jose77} have shown that for $p>4$, $h_p$ is an irrelevant scaling field down to intermediate temperatures, with the result that fluctuations restore the continuous symmetry of the 2dXY model above a threshold temperature, leading to a KTB transition~\cite{Kosterlitz73} and quasi-long range order over a finite range of temperature. Recently it has been shown~\cite{Wexler06} that a similar scenario remains valid even for infinitely strong crystal field strength, with the result that fluctuations restore continuous symmetry for $p$-state clock models with $p>4$, although for $4<p\le 6$ this occurs above the KTB temperature, $T_{\mathrm{KT}}$. For $p=2$ and $3$, $h_p$ is relevant, leading to phase transitions in the Ising and $3$-state Potts universality class respectively. $h_4$, on the other hand, is a marginal perturbation~\cite{Jose77}. A second order phase transition is predicted with non-universal critical exponents depending on the field strength. As $h_4 \rightarrow \infty$, XY$h_4$ crosses over to the $4$-state clock model, which is equivalent to two perpendicular Ising models, and the transition falls into the Ising universality class~\cite{Betts64}. The non-universal transition for XY$h_4$ is hence bounded by the Ising universality class for large $h_4$.

%The physical origin stems from the fact that, for $p>4$, thermal fluctuations can take spins from one crystal field minimum to another without destroying the local ordering. For $p=4$ the required jumps, $\Delta \theta = \pi/4$ are the same as those required, on average, to create a vortex around a square plaquette. Hence, the temperature at which the spins jump the crystal field barriers is the same as the KTB transition temperature.

The non-universal exponents of XY$h_4$ can be calculated analytically within the framework proposed by JKKN. They showed that to describe the evolution of the KTB transition in the presence of a weak $p$-fold field it is sufficient to replace (\ref{2DXY}) by the generalized Villain Hamiltonian~\cite{Jose77,Villain75}
\begin{eqnarray}\label{Vill}
  \frac{\mathcal{H}}{k_BT} = &-&K\sum_{\langle i,j\rangle} \left[1-{1\over{2}}(\theta_i-\theta_j-2\pi m_{ij})^2\right] +\sum_iipn_i \theta_i\nonumber\\
    &+&\log(y_0)\sum_i S_R^2 +\log(y_p)\sum_i n_i^2,
\end{eqnarray}
\noindent where $K=J/k_BT$. The integers $m_{ij}$ maintain the periodicity of the original Hamiltonian, for rotations over an angle $2\pi$. $S_R$ is  a directed sum of integers $m_{ij}$ around a square plaquette of four sites centred at $\vec R$: $S_R=m_{41}+m_{12}-m_{32}-m_{43}$, takes values, $S_R=0,\pm 1, \pm 2\dots$ and is therefore a quantum number for a vortex of spin circulation centred on the dual lattice site $\vec R$. $y_0$ is related to the chemical potential $\mu$ and fugacity $y$ for the creation of a vortex anti-vortex pair on neighbouring dual lattice sites: $y=y_0\exp(-\beta\mu)\approx y_0\exp(-\pi^2K/2)$. In the original Villain model $y_0=1$ but it is introduced here as a phenomenological small parameter which is renormalized in the subsequent flows. Similarly $y_p$ is a fugacity for a locking process of spins along one of the $p$-fold field directions with integer $n_i$ being a measure of the strength of this process at site $i$. For weak crystal fields, $y_p= {1\over{2}}{\tilde{h}}_p$ with $\tilde{h}_p=(h_p/k_BT)$, which reproduces the field contribution to the partition function to leading order in $y_p$. For strong fields $y_p \rightarrow 1$ and (\ref{Vill}) transforms into a discrete $p$-state model. Note however that this is not the $p$-state clock model: although the Villain model maintains the global rotational symmetry it destroys the local $O(2)$ symmetry of the pair interaction. The discrete terms $(\theta_i-\theta_j-2\pi m_{ij})^2$, $\theta_i = (n/p)2\pi,\; n=0,1,\dots, p-1$ hence do not have this symmetry over the interval $-\pi < (\theta_i-\theta_j-2\pi m_{ij}) < \pi$. For $p=4$ this means that neighbouring spins orientated perpendicularly have an energy less than half that of antiparallel spins and the ordered state has lower lying excitations than the corresponding clock model. It is therefore not clear whether the Villain model falls into the correct universality class in the strong field limit and for quantitative studies one should use Hamiltonian (\ref{2DXY}) rather than (\ref{Vill}).

With $y_p$ set equal to zero, a direct space renormalization analysis for the spin-spin correlation functions resulting from (\ref{Vill}) leads to RG flow equations for an effective coupling constant $K_{\mathrm{eff}}$ and vortex fugacity, $y$. For $K_{\mathrm{eff}}=2/\pi$, $y=0$, the flows yield the KTB transition~\cite{Kosterlitz74}. In the presence of the $p$-fold field the flow equations are modified and a third equation is generated \cite{Jose77,Jose77erratum}. For the explicit case with $p=4$, these are
\numparts
\begin{eqnarray}\label{RGEQ}
\left(K^{-1}\right)' &=& K^{-1} + 4\left(\pi^3y_0^2 e^{-\pi^2K}-\ 4\pi K^{-2}y_4^2 e^{-4K^{-1}}\right)\ln(b) \\
y_0' &=& y_0+(2-\pi K)y_0 \ln(b) \\
y_4' &=& y_4+\left(2-{4K^{-1}\over{\pi}}\right)y_4 \ln(b),
\end{eqnarray}
\endnumparts
where $b$ is the scale factor and where the equations are valid as $b \rightarrow 1$. This set of equations has fixed points at $K^{\ast}=2/\pi$, $y_0^{\ast}=\pm y_4^{\ast}$. We can calculate the linearized transformation matrix evaluated at the fixed point, $\ast$: $M_{i,j}={\partial K_i\over{\partial K_j}}|_{\ast}$, where $K_i=K^{-1},y_0,y_4$.

Solving for the eigenvalues we find
\begin{equation}
  \lambda=1,1+{\alpha\over{2}}\pm {1\over{2}}\sqrt{4a^2+\alpha^2},
\end{equation}
where $\alpha=16\pi^2(2\pi-1){\tilde{y}}^2 e^{-2\pi}\ln(b)$, $a^2=2\gamma\delta$, $\delta={4\over{\pi}}{\tilde{y}}\ln(b)$, $\gamma=8\pi^3{\tilde{y}} e^{-2\pi} \ln(b)$, and where $y_0=y_4={\tilde{y}}$. Writing $\lambda=b^{\sigma}$ we extract the three scaling exponents. There is one relevant exponent, which is interpreted as $\sigma_1=1/\nu$, the exponent taking the coupling constant away from the critical value at the now regular second order phase transition. There is also one irrelevant variable $\sigma_2$, which is interpreted as driving the vortex fugacity to zero. Finally, there is one marginal variable, $\sigma_3$, which, as announced, corresponds to the scaling exponent of the $4$-fold crystal field. Taking $h_4=0$ all eigenvalues become marginal, consistent with the particular scaling properties of the 2dXY model. In the small field limit, $\sigma_1=-\sigma_3=4\pi  e^{-\pi}\tilde{h}_4$ and $\sigma_2=0$. This gives the non-universal correlation length exponent~\cite{Jose77} $\nu \approx 1.8(k_BT_{\mathrm{KT}}/h_4)$. The strong field limit, $y_4=1$ gives $\nu \approx 0.47$, which should be compared with the exact result for the Ising model, $\nu=1$. The agreement is poor, as might be expected given the distortion of the four fold interaction imposed by the Villain model. It is clear from this result that a quantitative calculation for the strong field limit requires a different starting Hamiltonian.

In order to calculate $\beta$ from the scaling relations~\cite{Plischke94}, a second relevant scaling exponent is required. In this case the anomalous dimension exponent $\eta$ can be calculated directly from the correlation function~\cite{Jose77}. At the KTB transition of the XY model, $\eta=1/4$, giving the universal jump in the effective spin stiffness, $K_{\rm eff}=2/\pi$. It follows from the scaling relation $2\beta=(d-2+\eta)\nu$ that the finite size scaling exponent $\beta/\nu=1/8$, as in the Ising model, despite the fact that here the true $\beta$ and $\nu$ are not defined. This is an example of ``weak universality'' \cite{Suzuki74} between the two models. A striking result in the presence of a $4$-fold field is that $\eta$ remains unchanged to lowest order in $h_4$~\cite{Jose77}, indicating that a weak universal line extends out from the XY model along the $h_4$ axis. Here we make the hypothesis that the line extends right to the Ising limit, in which case $\eta=1/4$ for all $h_4$. This is clearly a reasonable assumption for the level of calculation made here. It is also an appealing result as other examples of weak universality are far less accessible to experiment \cite{Baxter82}. Analysis of the numerical data presented in the next section lends weight to this hypothesis, although the observed behaviour is found to divide into two regimes, depending on the strength of the $h_4$ field.

From this analysis we therefore predict a range of non-universal magnetization exponents going from
\begin{equation}\label{b-value}
  \beta\approx {1\over{8}}\left({1.8k_BT_{\mathrm{KT}}\over{h_4}}\right)
\end{equation}
for weak field, to $\beta=1/8$ in the strong field limit. To make quantitative comparison with simulation and experiment we need to estimate $\beta$ as a function of $h_4/J$. The critical value $K^{\ast}=2/\pi$ corresponds to a renormalized coupling constant, $J_{\mathrm{eff}}$, valid at large length scale such that $k_BT_{\mathrm{KT}}=\pi J_{\mathrm{eff}}/2$. In general $J_{\mathrm{eff}} < J$: for the Villain model $k_BT_{\mathrm{KT}}\approx 1.35 J$ \cite{Janke91}, while for the XY Hamiltonian (\ref{2DXY})  $k_BT_{\mathrm{KT}}/J \approx 0.9$ and is different again for more realistic Hamiltonians. Hence, while  we can make a theoretical prediction for the low field behaviour,
\begin{equation}\label{beta-scale}
\beta = {1\over{8}}\left({\alpha J\over{h_4}}\right),
\end{equation}
with $\alpha$ a constant of order unity, scaling equation (\ref{b-value}) by a factor $k_BT_{\mathrm{KT}}/J$ will probably not lead to an accurate quantitative estimate for $\alpha$ and the precise value is beyond the scope of the present calculation.

\section{Competition with Essential Finite Size Effects}

The survey of the $\beta$ values illustrated in Figure \ref{histogram} shows a clear discrepancy between theory presented above and experiment: the large values of $\beta$ predicted for small $h_4$ do not appear and the range of values is cut off at $\beta \approx 0.23$. As the latter is an effective exponent characteristic of XY criticality up to a finite length scale, it seems clear  that the non-universal critical phenomena are suppressed, for weak field, by the exceptional finite size scaling properties of the pure 2dXY model~\cite{Bramwell93,Chung99,Cuccoli03}. This hypothesis can be tested by numerical simulation, in which both $h_4$ and the system size may be directly controlled.

In a real XY system the relevant length scale will in most cases be less than the physical size of the system: for example, it could be a coherence length controlled by defects or dipolar  interactions \cite{Zhou07}, or, in the case of layered systems, a crossover scale to the third dimension~\cite{Hikami80,Bramwell93}. Thus, although real systems might have, for example, $10^{16}$ spins, the relevant scale for XY critical behaviour will typically be much smaller and compatible with the scale of Monte Carlo simulations, where the appropriate length scale is simply the system size. This finite length scale gives rise to a finite magnetization that disappears at the rounded KTB transition. As emphasised in~\cite{Bramwell93}, this is perfectly consistent with the Mermin-Wagner theorem~\cite{Mermin66}, which proves that the magnetization will be strictly zero in the thermodynamic limit. It is easy to convince oneself that finite size corrections to the thermodynamic limit are important for any physically realizable cut off length scale. The resulting low temperature magnetization is therefore directly relevant for experiment.

\begin{figure}%[htp]
     \centering
     \subfigure[]{
          \label{MvT}
          \includegraphics[scale=0.3]{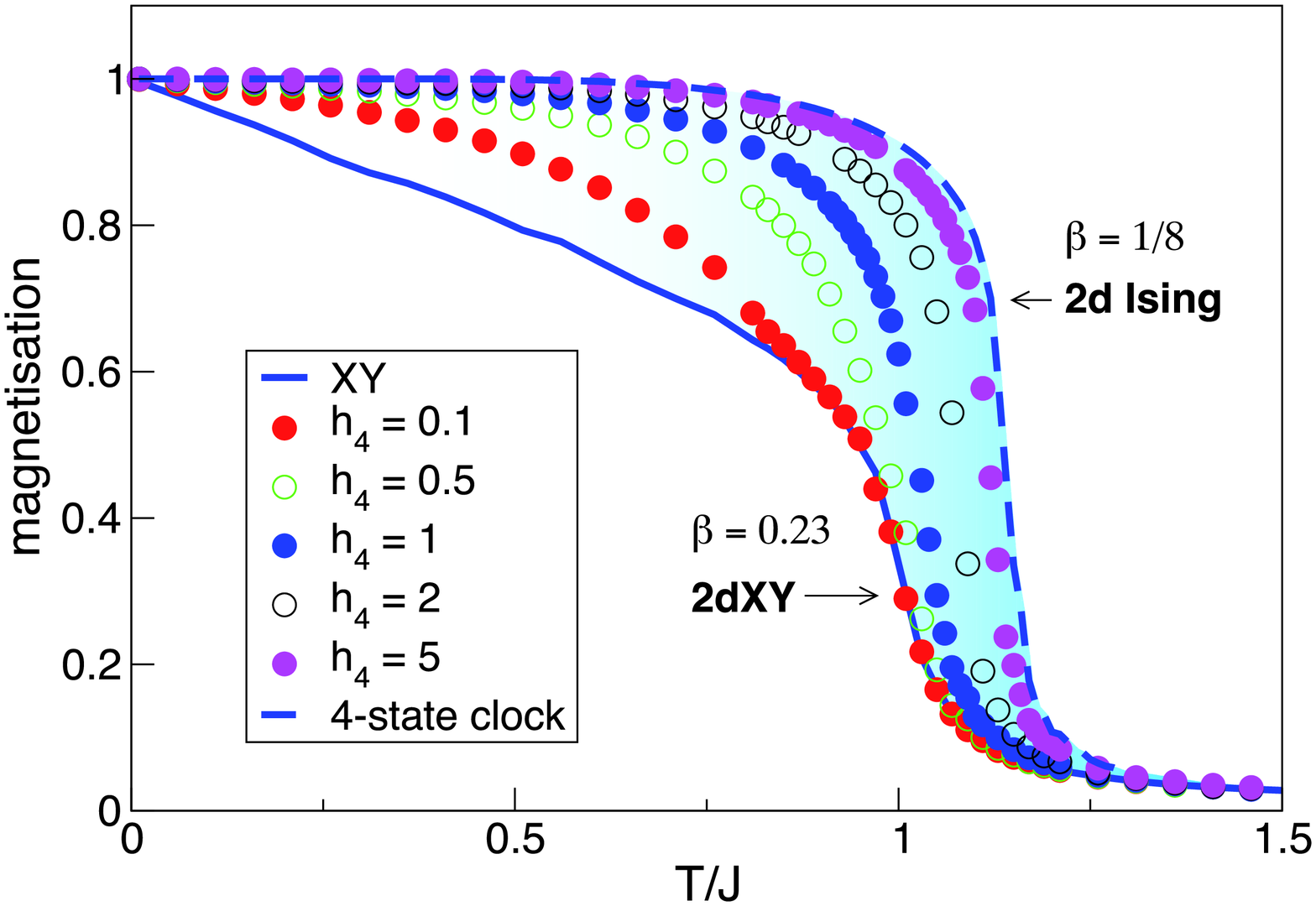}}
     \subfigure[]{
          \label{LogMvLogT}
          \includegraphics[scale=0.3]{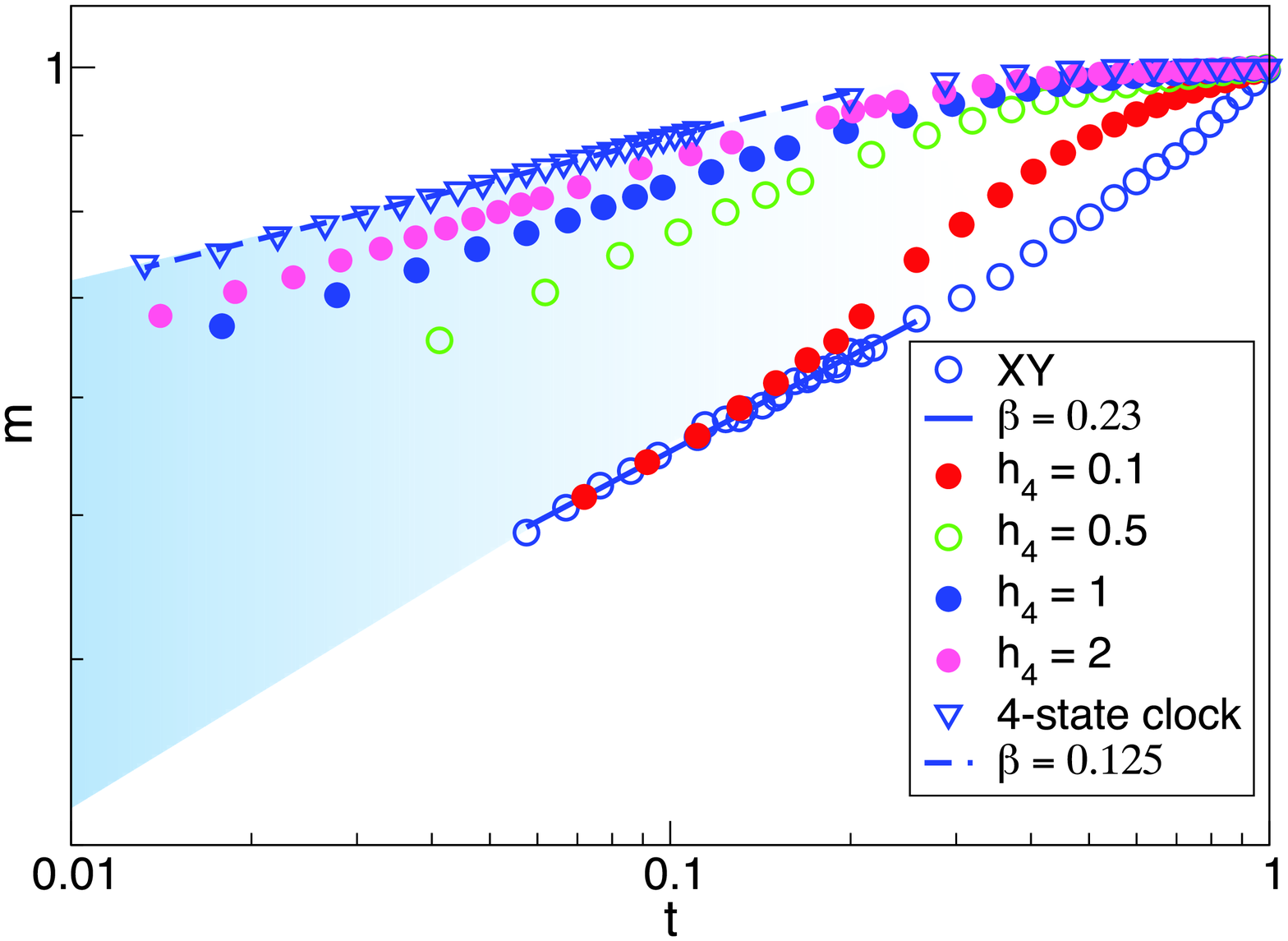}}
     \caption{Monte Carlo data for the 2dXY model in the presence of a 4-fold crystal field, with $N=10^4$. Plot (a) displays the magnetization data of the XY model in the presence of increasingly strong anisotropies. Plot (b) displays the same data as a function of reduced temperature $t$, in a logarithmic scale. In both plots the ``universal window'' is highlighted by the blue shading.}
  \label{XYstrong_h4_mag}
\end{figure}

In Figure \ref{MvT} we show the magnetic order parameter, the thermally averaged magnetic moment normalized to unity, versus temperature, with different $4$-fold field perturbations, for a system with $N=10^4$ spins. For $h_4=0$ the magnetization is characterized by the effective critical exponent, $\tilde{\beta} \approx 0.23$. A finite size analysis of Kosterlitz' renormalization group equations shows that in the region of the transition it approaches a universal number $\tilde{\beta}=3\pi^2/128\approx 0.23$, in agreement with both experiment and simulation data, such as that shown here. For weak crystal field, $h_4$ there is no change in the the region of the transition and the magnetization data coincide with the data for zero field~\cite{Bramwell97}. Only for $h_4/J \ge 0.5$ do they leave the zero field data through the transition, approaching results for the $4$-state clock model for large values of $h_4/J$. In Figure \ref{LogMvLogT} we show $\log(m)$ against $\log(t)$, where $t=(T-T_{\mathrm{c}})/ T_{\mathrm{c}}$. The transition temperature $T_{\mathrm{c}}$ is calculated from a finite size scaling analysis of the fourth order Binder cumulant for $M$ \cite{Binder81,Rastelli04-2,Zhou07} and is an estimate of the value in the thermodynamic limit. The slopes, for small $t$, give a first estimate of the exponent $\beta$, indicating that it lies in the interval $1/8 < \beta(h_4) < 0.23$ for all values of $h_4$, exactly as observed in experiment. The crossover to Ising behaviour is slow: for $h_4/J=1$, $\beta(h_4)\approx 0.15$ and to approach $\beta \approx 1/8$ requires $h_4/J$ in excess of $5$.

\begin{figure}%[ht]
     \centering
     \subfigure[]{
          \label{collapse-h1-1}
          \includegraphics[scale=0.25]{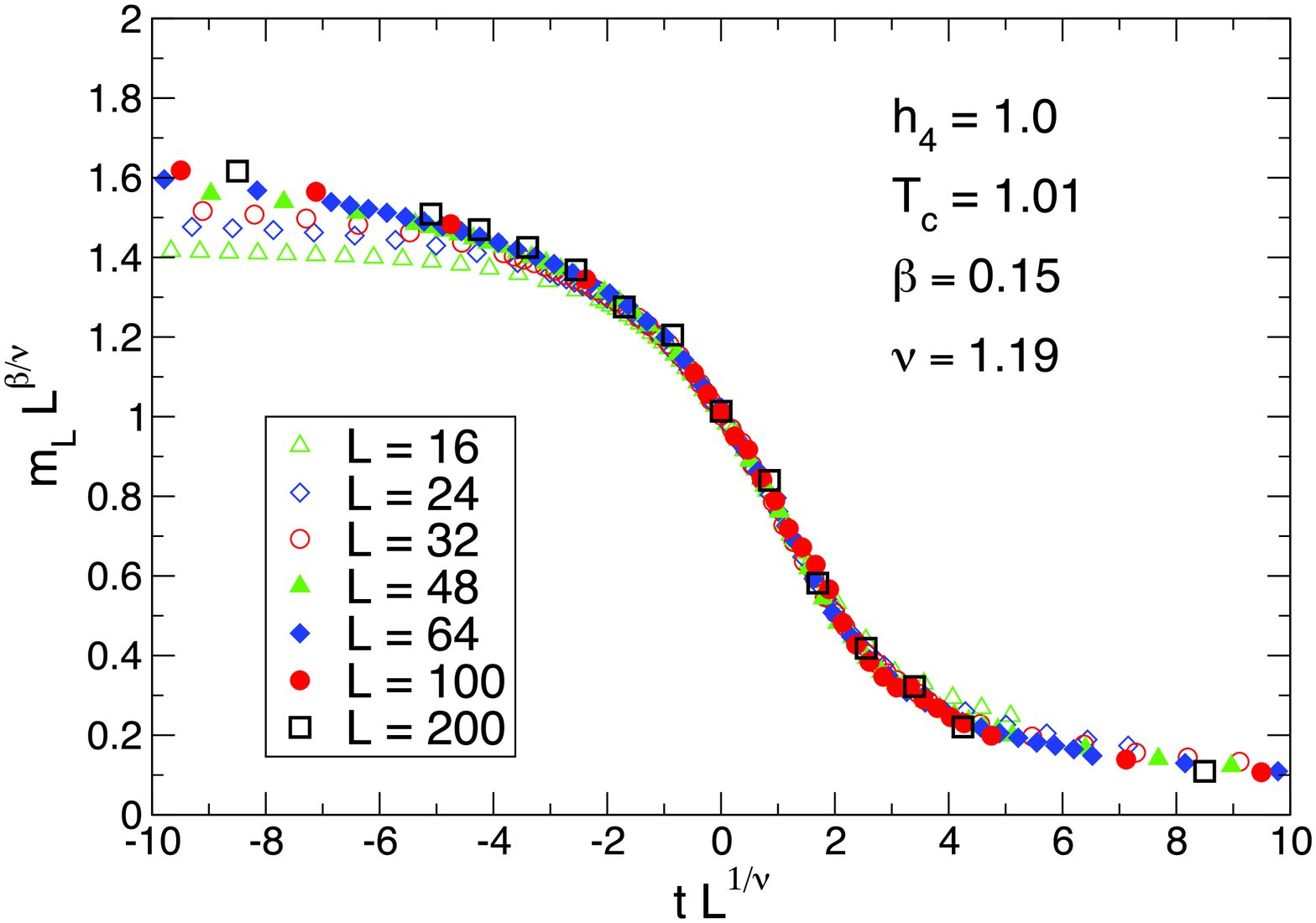}}
     \subfigure[]{
          \label{collapse-h1-2}
          \includegraphics[scale=0.25]{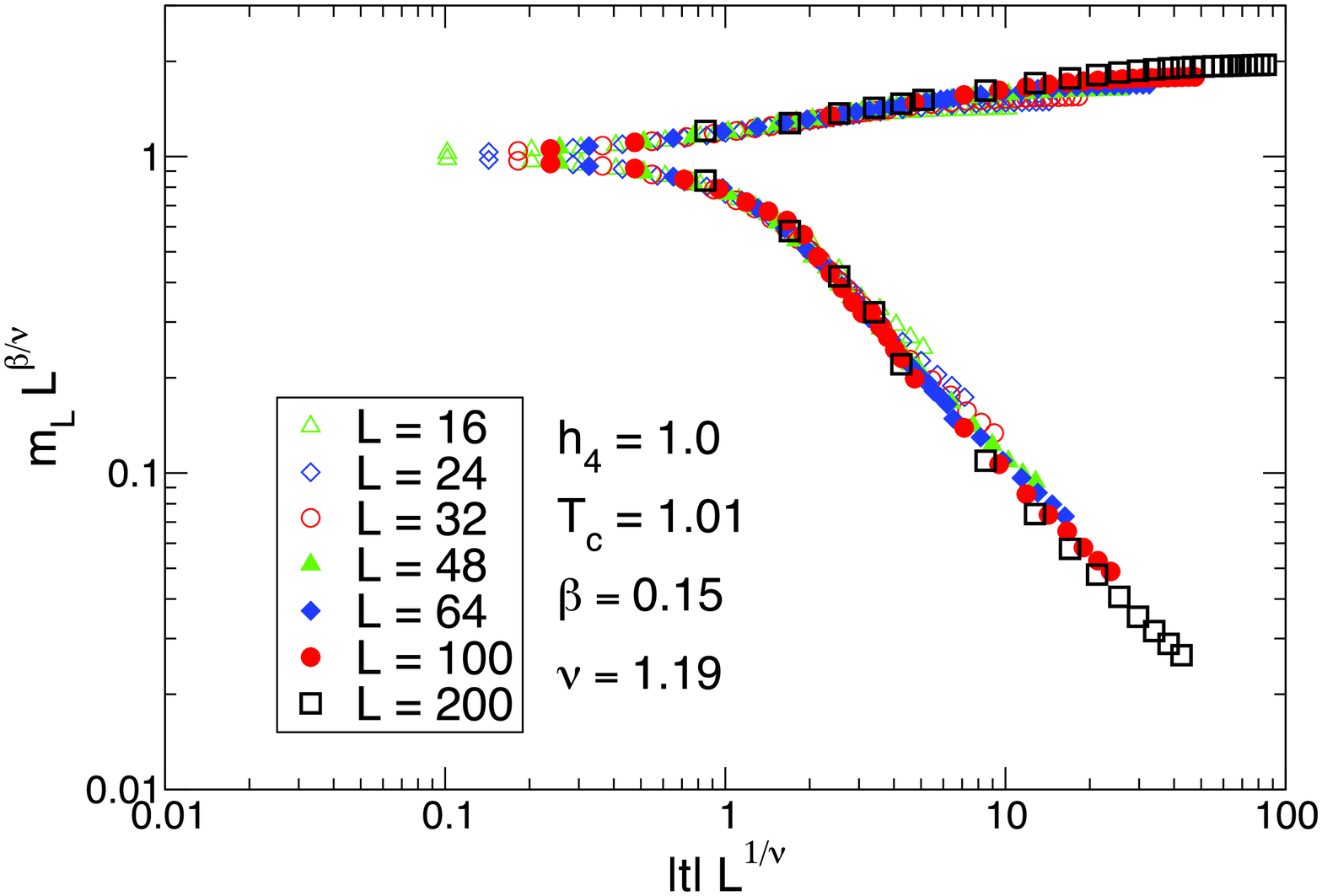}}
     \subfigure[]{
          \label{collapse-h2-1}
          \includegraphics[scale=0.25]{figure3c}}
     \subfigure[]{
          \label{collapse-h2-2}
          \includegraphics[scale=0.25]{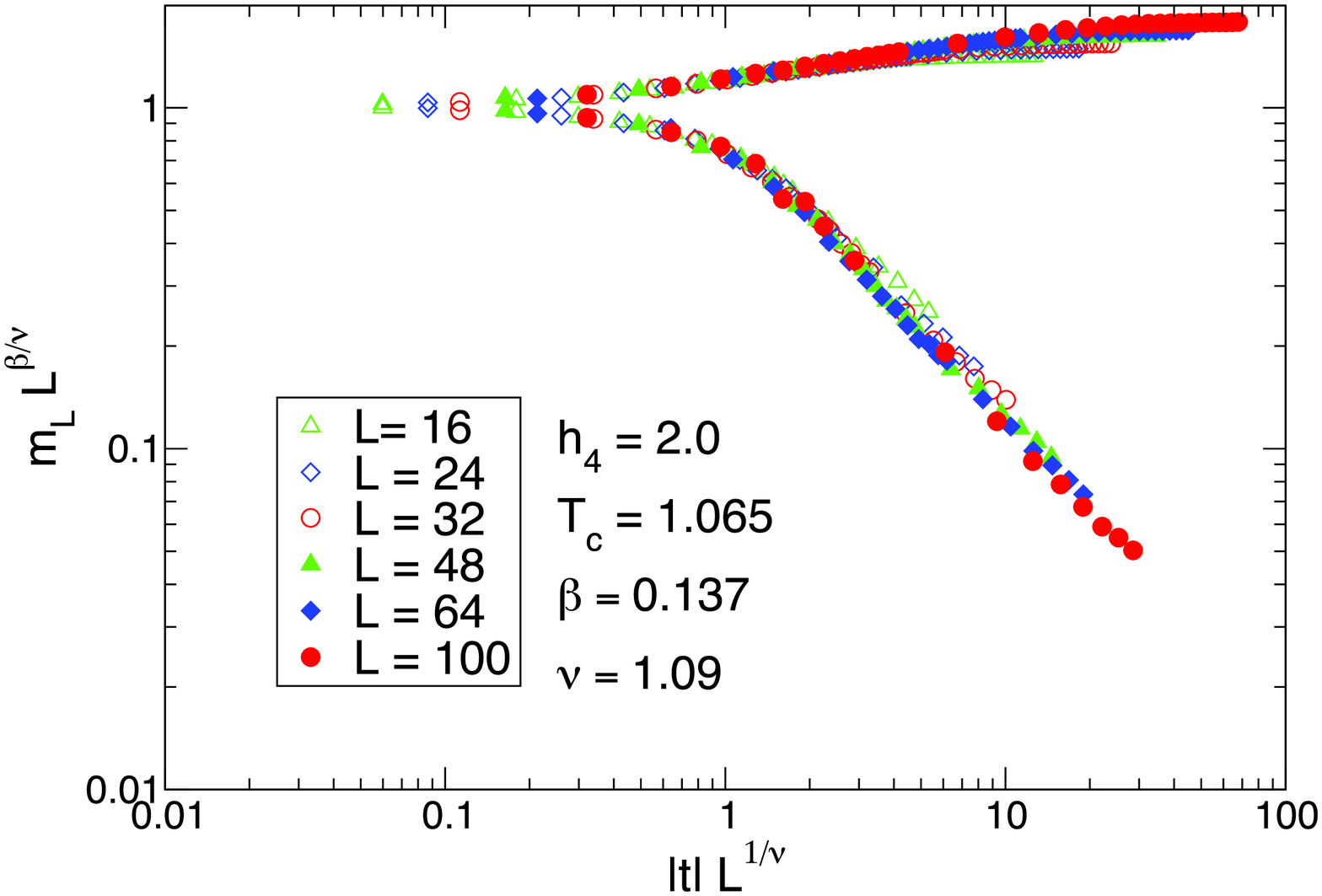}}
     \caption{Best data collapses for the 2dXY model with 4-fold crystal field $h_4/J=1.0$ and $h_4/J=2.0$ for different system sizes.}
  \label{collapse-h1-h2}
\end{figure}

Hence the data here, as in previous numerical work \cite{Bramwell97,Rastelli04,Rastelli04-2}, show evidence for a finite pocket of XY critical behaviour for small values of $h_4$. This appears to refute the prediction of JKKN, derived explicitly in the previous section, that the exponents vary continuously with $h_4$ \cite{Rastelli04} (see the further discussion below). For intermediate field strengths, however, the non-universal criticality does appear to hold as can be confirmed by a more detailed finite size scaling analysis. The values of $\beta$ and $\nu$ can be estimated more accurately by collapsing data for various system sizes onto the scaling relation $ML^{\beta/\nu}\sim tL^{1/\nu}$. The best data collapses for $h_4/J=1$ and $h_4/J=2$, with $T_{\mathrm{c}}$ in each case fixed from the Binder cumulant calculation, are shown in Figure \ref{collapse-h1-h2}. We find $\beta=0.148(5)$ and $\beta=0.136(10)$, in good agreement with the values found from Figure \ref{LogMvLogT}, and $\nu=1.19(4)$, $\nu=1.09(8)$. The ratio $\beta/\nu=0.126(4)$ in each case, is in agreement with the weak universality hypothesis. Similar results for $h_4/J=0.5$ can be found in \cite{Rastelli04}. Although these exponent values are not so different from the Ising model values, the data collapse is less satisfactory when Ising exponents are used.

\begin{figure}%[ht]
   \centering
   \includegraphics[scale=0.3]{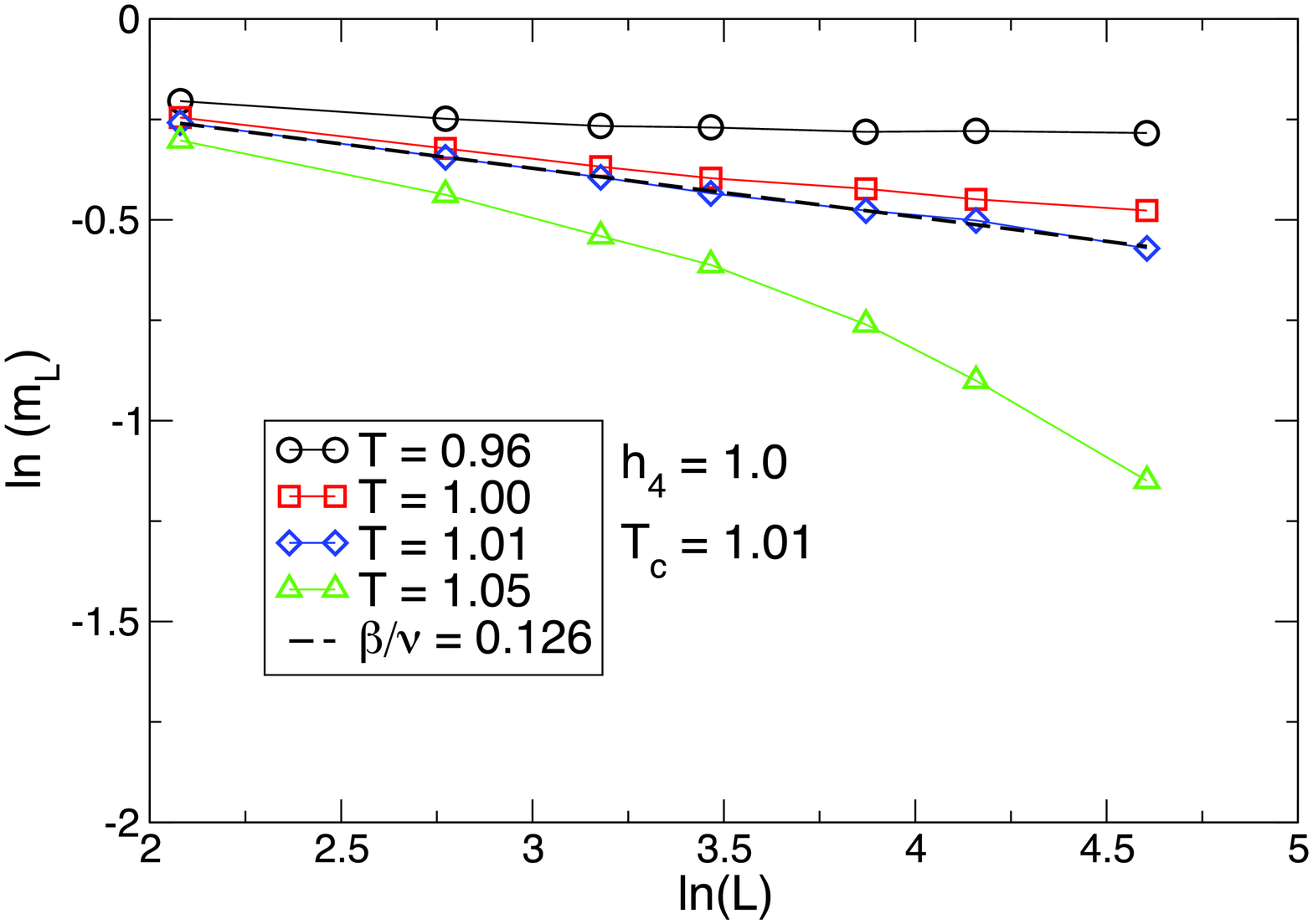}
   \caption{Magnetisation $m$ against system size $L$, in log-log scale, for the 2dXY model with 4-fold crystal field $h_4/J=1.0$, for temperatures near the transition.}
   \label{FSS-H1}
\end{figure}

Further evidence for weak universality at intermediate field strengths can be found from studying the finite size scaling properties of $m$ at the transition. In Figure \ref{FSS-H1} we show $\log(m)$ against $\log(L)$ for $h_4/J=1$ for a range of temperatures near the transition. At the transition one expects a power law evolution characterised by the finite size scaling exponent $\eta/2={\beta/\nu}$. The best power law occurs at $T_{\mathrm{c}}=1.010(5) J$, which is the same as the value found from the Binder cumulant method. The scaling exponent $\eta/2=0.126(3)$, is the same as that found for the data collapse in Figures \ref{collapse-h1-1} and \ref{collapse-h1-2}.

This and previous numerical work~\cite{Bramwell97,Rastelli04,Rastelli04-2,Schneider76} are consistent with $h_4$ being marginal. In this case the crossover exponent to the new universality class is zero so crossover occurs, at best on exponentially large length scales, as a result of corrections to scaling~\cite{Bramwell97}. Hence, for small and intermediate crystal field strengths the finite size scaling appears compatible with that of the continuous symmetry of the 2dXY model~\cite{Bramwell97}, as in the $6$-fold case. In fact the most detailed finite size scaling analysis~\cite{Rastelli04} shows no evidence of such a crossover for small fields. It therefore remains an open question whether the pocket of pure XY behaviour for small $h_4/J$ is a pragmatic observation related to excessively slow crossover, or whether it remains right to the thermodynamic limit. In either case this is the main result of this section: large values of $\beta$ are indeed masked by the pocket of 2dXY behaviour, leading to the effective exponent $\tilde{\beta}$ for weak $h_4$ and creating a divide between systems with strong and weak $4$-fold fields, with the non-universal character of XY$h_4$ only appearing for $\beta(h_4) < 0.23$. The threshold value  of $h_4$, separating the two regimes  can be estimated theoretically by putting $\beta(h_4)=0.23$ in Equation (\ref{beta-scale}). Using $k_BT_{\mathrm{KT}}/J \approx 0.9$  gives $\alpha=1.6$ and  $h_4/J \approx 0.9$, a ratio of order unity, in agreement with the above general arguments, but an over estimate compared with numerics, where the change of regime occurs for $h_4/J\sim 0.5$, corresponding to $\alpha\approx 1$.

\begin{figure}
   \centering
   \includegraphics[scale=0.3]{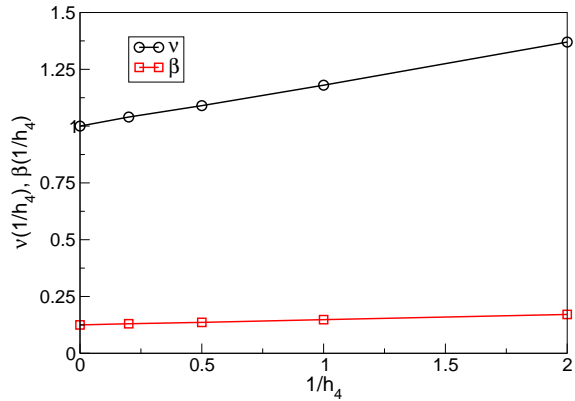}
   \caption{Exponents $\beta$ and $\nu$ measured from a finite size scaling analysis of the Monte Carlo data plotted against $1/h_4$.}
   \label{betanu}
\end{figure}

Having confirmed that $\eta \approx 0.25$ over the whole range of $h_4$, we finally fix $\eta = 0.25$ and use our estimates of $\nu(h_4)$ from the scaling collapse to give a further estimate of the exponents as a function of $h_4$. The estimates of $\nu$ and $\beta$, summarised in Table \ref{exponents}, are in good agreement with all previous unconstrained estimates. We also include estimates of $\beta$ derived by a typical experimental analysis of fixing $T_{\mathrm{c}}$ from the maximum in the susceptibility or where the magnetization approaches zero, and deriving $\beta$ from a log-log plot. There is seen to be a systematic error between the different estimates of $\beta$, especially for small values of $h_4$. Nevertheless, the experimental exponents are still found to lie in the universal window of values predicted for the ``true'' exponents of the underlying model. The various critical exponents plotted  in Figure \ref{betanu} are found to be linear in  $1/h_4$. By fitting to $\beta(h_4) = 0.125+a/h_4$, we estimate the constant $a$ to be 0.032 for the ``true'' exponents, and 0.05 for the experimental exponents. These values are clearly very different from that expected for the constant $\alpha$ in Equation (\ref{beta-scale}), but once outside the pocket of pure XY behaviour we are no longer in the weak field regime for which  Equation (\ref{beta-scale}) is valid, as was shown in the previous section.

\begin{table}
\footnotesize
\caption{Critical exponents for the XY$h_4$ model, as determined from a
finite size scaling analysis, and as measured directly from Monte Carlo
magnetization data for a system of size $L=100$.}
\vspace{1mm}
\begin{indented}
\item[]\begin{tabular}{ccccc}
\hline
\hline
$h_4$ & $\nu$ & $\beta$ & $T_{\mathrm{c}}^L$ & $\beta(T_{\mathrm{c}}^L)$ \\
\hline
0.5      & 1.37(6) & 0.171(10) & 1.01(1) & 0.214(9) \\
1        & 1.19(4) & 0.148(5)  & 1.04(1) & 0.196(6) \\
2        & 1.09(8) & 0.136(10) & 1.08(1) & 0.155(3) \\
5        & 1.04(6) & 0.130(7)  & 1.12(1) & 0.129(3) \\
$\infty$ & 1.00(5) & 0.125(6)  & 1.14(1) & 0.123(3) \\
\hline
\hline
\end{tabular}
\end{indented}
\label{exponents}
\normalsize
\end{table}

\section{The Strong Field--Weak Field Divide}

Experimental evidence for the strong field--weak field divide comes from making head to head 
comparisons between systems listed in the appendix.

A quantitative comparison is afforded by the ferromagnet Rb$_2$CrCl$_4$~\cite{Hutchings81} and the 
antiferromagnet K$_2$FeF$_4$~\cite{Thurlings89}, both quasi two dimensional square lattice systems 
with $S = 2$. In both systems the intra-plane isotropic exchange coupling, $J$ is much bigger than the inter-plane value $J'$, giving rise to an extended temperature range with two dimensional critical fluctuations. However, while the ferromagnet shows all the characteristics of the pure XY universality class~\cite{Als-Nielsen93,Bramwell95}, the antiferromagnet has non-universal exponents, with $\beta=0.15$~\cite{Thurlings89}, which we now see to be consistent with XY$h_4$. A realistic model Hamiltonian for either system has the following form

\begin{equation}
 \mathcal{H} = J_0 \sum_{\langle i,j\rangle} \mathbf{S}_i \cdot \mathbf{S}_j + D \sum_i (S_i^z)^2 + \frac{1}{2} e (S_+^4 + S_-^4),
\end{equation}\label{H-KFeF}

\noindent where the weak interplane  exchange and a weak fourth order axial term are ignored (in the 
case of Rb$_2$CrCl$_4$ small departures from tetragonal symmetry are neglected for the purpose of 
this discussion).

The crystal field $D$ confines the spins to an easy plane breaking the $O(3)$ rotational symmetry of the Heisenberg exchange and the 4-fold term $e$ breaks symmetry within that plane, putting Hamiltonian (\ref{H-KFeF}) in the XY$h_4$  universality class. For both Rb$_2$CrCl$_4$ and K$_2$FeF$_4$ accurate estimates of the Hamiltonian parameters were derived by fitting magnon dispersions measured by neutron scattering to a self-consistent spin wave calculation \cite{Thurlings89,Hutchings81,Oguchi60}. However, in order to fit the spectra, the fourth order term was decoupled into an effective second order term, with amplitude $E \approx 6eS^2$. In the low temperature limit one can estimate parameters $\tilde{e} = \left| eS^4/J_0S^2\right|$ and $\tilde{D} = \left|DS^2/J_0S^2\right|$. Values are shown in Table \ref{layered_parameters} for both materials.

\begin{table}
\caption{Main parameters for K$_2$FeF$_4$ and Rb$_2$CrCl$_4$, as determined from experiment~
\cite{Thurlings89,Hutchings81}.}
\vspace{1mm}
\begin{indented}
\item[]\begin{tabular}{lcc}
\hline
\hline
& K$_2$FeF$_4$ & Rb$_2$CrCl$_4$ \\
\hline
$S$ & $2$ & $2$ \\
$J$($\kelvin$) & -15.7 & 15.12 \\
$D$($\kelvin$) & 5.7 & 1.06 \\
$E$($\kelvin$) & -0.49 & 0.123 \\
$\tilde{D}$ & 0.363 & 0.07 \\
$\tilde{e}$ & 0.0052 & 0.0013 \\
$\beta$ & 0.15(1) & 0.230(2) \\
\hline
\hline
\end{tabular}
\end{indented}
\label{layered_parameters}
\end{table}

To get an estimate of the 4-fold field that determines the critical exponents, it is tempting
to assume that systems with $S=2$ are classical and to associate $\tilde{e}$ with the parameter $h_4/J$ arising from Equation (\ref{2DXY}). The parameter $\tilde{e} = 0.0013$ for Rb$_2$CrCl$_4$ and $0.0052$ for K$_2$FeF$_4$, which seem sufficiently small to put both systems into the weak field regime with pure XY universality, as has been directly confirmed by numerical simulation~\cite{Taroni07}. However, assigning an effective classical Hamiltonian of the form (\ref{2DXY}) for systems with finite $S$ is not so straightforward: for finite $S$, through the  uncertainty principle, out-of-plane and in-plane spin fluctuations are not statistically independent. As a consequence the energy scale for in-plane spin rotations and the consequent effective value for $h_4$ depend collectively on $J$ and $D$ as well as on $\tilde{e}$. This can be seen from a detailed consideration of the magnon dispersion arising from (\ref{H-KFeF}). This calculation reveals a distinct difference between the ferromagnetic and antiferromagnetic cases, with the latter retaining strong quantum effects even for $S=2$. For the antiferromagnet one finds two magnon branches which, for $D=e=0$, are degenerate and gapless for zero wavevector and where each mode constitutes a conjugate in-plane and out-of-plane spin fluctuation term of equal amplitude. For finite crystal field strength the degeneracy is lifted, energy gaps appear everywhere in the spectrum and the symmetry is broken between the in-plane and out-of-plane fluctuation amplitudes. In the following we refer to a mode as in-plane or out-of-plane if the conjugate variable with the largest amplitude is in, or out of the plane. To lowest order in $1/S$ the out-of-plane branch develops a gap at zero wave vector:
\begin{equation}
 \Delta_1 =S\left[2(D+|E|)(2|J|z+4|E|)\right]^{1/2} \approx 2S\sqrt{Dz|J|},
 \end{equation}\label{delta1}
while the in-plane branch has 
\begin{equation}
\Delta_2 =S\left[(4|E|)(2|J|z+2D+2|E|)\right]^{1/2}\approx 2S\sqrt{2|E|z|J|},
\end{equation}\label{delta2}
(here $z$ = 4). These gaps depend on the geometric mean of the exchange field $zJ$ and the crystal field $e$ or $D$ with the result that they are surprisingly large on the scale of $J$, as noticed by Thurlings {\it et al.}~\cite{Thurlings89}. For K$_2$FeF$_4$  $\Delta_1=70.8$ K, $\Delta_2=23.9$ K at $4.2$ K, renormalising only weakly with temperature~\cite{Thurlings89}. $\Delta_1$ is larger than the transition temperature, $T_\mathrm{N}=63\; K\approx JS^2$, so the out-of-plane branch of spin fluctuations will be frozen by quantum effects over the whole of the ordered phase, leaving the predominantly in-plane spin fluctuations only. Interpreting these as the classical fluctuations in an effective plane rotator model with Hamiltonian (\ref{2DXY}) leads to a crystal field, $h_4(\mathrm{eff})$, of the order of $\Delta_2$. This gives $h_4(\mathrm{eff})/JS^2 \sim 0.33$ which is the right order of magnitude to fall into the strong field category. The fact that for K$_2$FeF$_4$, $T_\mathrm{N}/ JS^2\approx 1$, as is the case for the model systems with Hamiltonian (\ref{2DXY}) presented in the previous section, is highly consistent with this interpretation. Higher order terms in $1/S$ renormalize $D$ and $|E|$ such that the values given in table \ref{layered_parameters} are higher than those predicted by fitting to linear spin wave theory. One can further speculate that quantum fluctuations for the in-plane branch will renormalize the effective $h_4$ in (\ref{2DXY})~\cite{Rastelli04-2} to an even higher value. The non-universal exponents observed for K$_2$FeF$_4$ could therefore be examples of the XY$h_4$ universality class.

For the ferromagnet Rb$_2$CrCl$_4$, $D$ flattens the cone of spin precession giving a range of $q$ values where the energy spectrum varies approximately linearly with wave vector, but does not open a gap. The field $e$ opens a zero wave vector energy gap that varies as $\sqrt{De}$. It is of order $1$ K, decreasing to zero at the transition temperature $T_{\mathrm{c}} = 52$ K, and so can hardly affect the thermodynamics in the critical region. Although the effective value of $h_4$ depends on the geometric mean of $D$ and $e$ rather than just the bare value of $e$ it is independent of $J$ and hence much smaller than than for the antiferromagnetic case. This places Rb$_2$CrCl$_4$ in the weak field regime, consistent with the observation of XY universality for this material~\cite{Als-Nielsen93, Bramwell95}.

From this comparison, it seems likely that magnetic systems that show true XY$h_4$ universality will mostly be antiferromagnetic. Indeed a similar ``spin dimensional reduction'' due to quantum
suppression of fluctuations has recently been observed in quantum Monte Carlo simulations with a
Hamiltonian similar to (\ref{H-KFeF})~\cite{Cuccoli03}. More calculations beyond the spin wave approximation are required to clarify this point.

Among non-magnetic systems, oxygen absorbed onto Mo(110)~\cite{Grzelakowski90} or W(110)~\cite
{Baek93} and hydrogen on W(110)~\cite{Lyuksyutov81} have both been claimed to fall into the XY$h_4
$ class, representing four fold equivalents of the two stage melting process for hexagonally coordinated systems~\cite{Nelson79}. Note that the (110) surface does not have four fold symmetry, but if we adopt these claims as a premise, then a comparison of the two systems is indeed perfectly consistent with XY $h_4$ universality and with the preceding arguments about the strong field-weak field divide. Electron hybridization between absorbed and substrate particles will result in the generation of electronic dipoles aligned perpendicularly to the (110) surface. The resulting $1/r^3$ interaction between the particles is repulsive and of sufficiently long range to ensure crystalization into a square lattice. The (110) surface provides a substrate potential with 4-fold topology (though not 4-fold symmetry) and which can be made commensurate with the free standing array by tuning the adsorbate density, the clearest example being the $(2\times 2)$ lattice structure~\cite{Lyuksyutov81}. The result is claimed to be in the XY$h_4$ universality class and the measured exponents, $\beta\simeq 0.19$~\cite{Grzelakowski90,Baek93}, are, in light of the current work, consistent with this. In principle, the same should be true for the $(2\times 2)$ ordering transition for hydrogen on W(110) but the measured $\beta$, 0.25, is consistent with the pure XY model~\cite{Lyuksyutov81}. Hydrogen being so much lighter that oxygen, larger zero point fluctuations 
should make the substrate potential less effective at pinning the crystal, putting it in the category of systems with a weak field $h_4$, consistent with the experimental observation.

\section{Other Exponents and Scaling Relations}\label{other_exponents}

Further evidence for the experimental relevance of the finite size effects is found in the behaviour of other critical exponents. The exponent $\eta$, which according to the previous weak universality arguments should be $0.25$ for all $h_4$, is only found to closely approximate the theoretical value for model Ising systems such as Rb$_2$CoF$_4$~\cite{Ikeda79}. For model XY systems the predicted $\eta = 0.25$ or $\delta=15$ are always observed at a temperature {\it well below} $T_{\mathrm{c}}$ (say $0.9$ $T_{\mathrm{c}}$), with $\eta(T)$ increasing to larger values at $T_c$ and $\delta(T)$ decreasing, since $\delta = (4-\eta)/\eta$. For example in the XY layered ferromagnets Rb$_2$CrCl$_4$ and K$_2$CuF$_4$ $\eta(T)$ and $\delta(T)$ have been measured with precision by several different methods~\cite{Hirakawa82,Cornelius86,Bramwell95}: in both cases $\eta$ rises to about $\eta=0.35$ at $T_{\mathrm{c}}$. This is a very strong signature of the finite size scaling properties of the XY model and is consistent with the predicted logarithmic shift in transition temperature, $\left[T_{\mathrm{c}}(L)-T_{KT}\right]\sim {1\over{\log^2(L)}}$~\cite{Bramwell93,Kosterlitz74}, for a finite size system~\cite{Cardy96}. As the measured value of $\eta$ increases continuously through the transition its value at $T_{\mathrm{c}}(L)$ is thus expected to be in excess of $\eta=1/4$.

It seems that the anomalous value of $\eta > 0.25$ extends to systems with XY$h_4$ universality: for example, in K$_2$FeF$_4$ it is estimated to be $\eta \approx 0.35$. This is again consistent with the shift in transition temperature observed in finite size systems. Defining $T_{\mathrm{c}}(L)$ from the maximum susceptibility leads to a shift, $\left[T_{\mathrm{c}}(L) - T_{\mathrm{c}}\right] \sim L^{-1/\nu}$. Here, in the four fold field problem, $\nu>1$ which means that shift remains important even in the intermediate field regime. Referring to Figure \ref{FSS-H1}, one can see that extracting a critical exponent from the initial slope, for $T>T_{\mathrm{c}}$, will lead to an overestimate of $\eta$. As experiments do not, in general, have access to the finite size scaling information available to numerical studies, it seems reasonable that the experimental $\eta$ values are generally larger than the expected thermodynamic limit value. Thus, we propose that $\eta(T_{\mathrm{c}})$ appearing greater that 1/4 remains a finite size effect.

Similarly, the measured values of $\nu$ are systematically smaller than unity, while a consequence of weak universality is that $\nu$ should be greater than one for all finite $h_4$. For example for K$_2$FeF$_4$ $\nu \approx 0.9$, giving $\beta/\nu = 0.16$, greater than the predicted ratio $1/8$, but together with $\gamma \approx 1.5$ the set of exponents do satisfy the hyperscaling relation, $2\beta+\gamma =d\nu$, as well as the relation $\beta/\nu = d-2 + \eta/2$. The same holds true for oxygen on W(110)~\cite{Baek93}, for which $\beta$ and $\gamma$ have been determined to be 0.19 and 1.48, respectively. The shift in $\nu$ is therefore consistent with the shift in $\eta$. It seems reasonable to assume that these changes are also due to finite size effects, which at present prevent the observation of the weak universal line we have shown evidence for, for all values of $h_4$. More detailed experimental and numerical studies to clarify this point would be of great interest.

\section{Conclusions}

In conclusion, the XY model with four fold crystal field is of relevance to a great number of experimental two dimensional systems. We have focused on the largest experimental data sets, those for two-dimensional magnets, adsorbed gaseous monolayers and in particular on the measured exponent $\beta$. With regard to the histogram in Figure \ref{histogram}, the systems that comprise it can only be fully understood on a case by case basis.

However, we show in this paper that the Hamiltonian (\ref{2DXY}) contains the principle two dimensional universality classes that are relevant to experiment and that a uniform distribution of values $h_4$ would, because of the marginal finite size scaling properties of the model, produce a probability density of the same form as Figure \ref{histogram} with a continuous spectrum bounded by peaks at the Ising and XY limits. This is what we refer to as the universal window for critical exponents. We have further shown that the actual values of the four fold crystal field that occur in real systems are, at first sight, too small to take any system away from the XY limit. However, we have identified at least one mechanism, in antiferromagnets, whereby the four fold field is effectively amplified by quantum confinement of the spins to the easy plane. Other mechanisms of realising XY$h_4$ universality are possible in individual cases~\cite{Prakash90,Landau85}. We have demonstrated the relevance of finite size scaling corrections to the experimental data set, with the relevant length scale giving a crossover away from XY criticality. Future work should focus on the finite size scaling aspects and on individual systems to see if a more accurate quantitative connection between the physical $h_4$ and the observed critical behaviour can be established. Further to this, we propose here that the non-universal exponents of XY$h_4$ should satisfy weak universality, with $\beta/\nu=1/8$ for all $h_4$ and we have given evidence that this is true in the range of intermediate field values. The robustness of the pocket of true XY behaviour, observed for weak fields~\cite{Bramwell97, Rastelli04}, in the thermodynamic limit remains an open question. Finally we remark that all evidence confirms that truly two dimensional systems, quasi two dimensional systems and numerical simulations reveal the same syndrome of behaviour, so much can be learned about new two dimensional systems~\cite{Rose07,Parnaste07} through comparisons with old ones~\cite{Thurlings89,Hutchings81}. It is fortunate that there is such an extensive and carefully determined data base.

\ack

It is a pleasure to thank Maxime Clusel, Martin Greven, Bj\"{o}rgvin Hj\"{o}rvarsson and Marco Picco for useful discussions. PCWH thanks the London Centre for Nanotechnology and the Royal Society for financial support and the Rudolph Peierls Institute for Theoretical Physics, University of Oxford, for hospitality during the completion of this work. AT thanks the EPSRC for a studentship.

\appendix
\section*{Appendix: Construction of the Histogram of $\beta$ Exponents}\label{appendix}
\setcounter{section}{1}

In constructing the histogram of experimental two dimensional $\beta$ exponents, a number of factors were considered. First, it was crucial to avoid circular logic by excluding those systems which were assigned a dimensionality purely on the basis of their exponent values, rather than on a large body of experimental evidence. Fortunately, we found no such cases in the literature. Therefore all systems included in the histogram are assigned as two dimensional on the basis of compelling experimental evidence of two dimensionality. Likewise we found no examples of systems considered to be two dimensional that exhibit $\beta \approx 1/3$, which might, in the absence of extra evidence, be mistakenly assigned as three dimensional systems and wrongly excluded from the data set. It should be noted that in layered magnets, the crossover from two dimensional to three dimensional exponents is generally very sharply defined so there is no ambiguity in identifying the two dimensional regime. A second criterion for inclusion in the histogram was that the exponents were determined with reasonable precision and accuracy (typically $\Delta \beta < \pm 0.01$). This inevitably necessitated a subjective judgement, but only a few results were excluded on these grounds: typically those exponents determined by powder (rather than single crystal) neutron diffraction, which is generally accepted to be inadequate for the accurate determination of $\beta$. The experimental exponents are generally not asymptotic exponents, but the numerical study presented above reveals that the difference between asymptotic exponents and those determined using finite size scaling techniques at temperatures down to $\sim 0.9$ $T_{\mathrm{c}}^L$ is generally negligible at the level of accuracy required for the present purpose. The histogram also excludes a number of interesting systems on the basis of there being legitimate grounds for alternative explanations for their observed critical behaviour. These include metamagnetic materials~\cite{Rujiwatra01}, systems undergoing spin-Peierls transitions~\cite{Gaulin00,Lorenzo99,Birgenau99}, and bulk systems undergoing order-disorder transitions~\cite{Harris99}.

The following tables lists all the systems included in the histogram. Table \ref{layered_table} contains data for layered magnets, and includes examples of molecular magnets~\cite{Blundell04}. Note that K$_2$MnF$_4$ represents two data points in the histogram as the elegant work of van de Kamp {\it et al.}~\cite{vandeKamp98} used a magnetic field to tune the system between Ising and XY symmetry, with $\beta$ recorded for both cases. In all other cases the $\beta$'s are determined in zero applied field. Table \ref{thinfilm_table} contains data for ultrathin magnetic films. Although there are several cases in which films of different thicknesses have been measured in order to study crossover to three dimensional behaviour, only the values of $\beta$ in the two dimensional limit are reported here, and are included as only one data point in the histogram. Finally, Table \ref{orderdisorder} includes data for adsorbed gaseous monolayers, and systems undergoing surface reconstruction and melting processes.

Our aim has been to make an exhaustive survey up to the time of publication. We apologise to any authors whose work we may have inadvertently overlooked, but we are confident that these cases would not significantly modify the form of the histogram.

\begin{table}
\footnotesize
\caption{List of 2d critical exponents $\beta$ for layered magnets reported in the literature, mostly measured by neutron diffraction (F = ferromagnet, A = antiferromagnet, Fo = (HCO$_2$), chdc = \emph{trans}-1,4-cyclohexanedicarboxylate, 5CAP = 2-amino-5-chloropyridinium).}
\vspace{1mm}
\begin{tabular}{lccccc}
\hline 
\hline
System & $\beta$ & $t$ range & $T_{\mathrm{c}}$ ($\kelvin$) & Type & Reference \\ \hline

Rb$_2$CoF$_4$ & 0.119(8) & 1$\cdot$10$^{-1}-$2$\cdot$10$^{-4}$ & 102.96 & A & \cite{Samuelsen73} \\
ErBa$_2$Cu$_3$O$_7$ & 0.122(4) & 0.11$-$1$\cdot$10$^{-4}$ & 0.618 & A & \cite{Lynn89} \\
K$_2$CoF$_4$ & 0.123(8) & 1$\cdot$10$^{-1}-$8$\cdot$10$^{-4}$ & 107.85 & A & \cite{Ikeda74} \\
BaNi$_2$(AsO$_4$)$_2$ & 0.135 & 3$\cdot$10$^{-1}-$1$\cdot$10$^{-2}$ & 19.2 & A & \cite{Regnault90} \\
Ba$_2$FeF$_6\,^{\ast}$ & 0.135(3) & 7$\cdot$10$^{-1}-$4$\cdot$10$^{-3}$ & 47.96(4) & A & \cite{Brennan93} \\
K$_2$NiF$_4$ & 0.138(4) & 2$\cdot$10$^{-1}-$3$\cdot$10$^{-4}$ & 97.23 & A & \cite{Birgenau70} \\
K$_3$Mn$_2$F$_7$ & 0.154(6) & 1$\cdot$10$^{-1}-$1$\cdot$10$^{-3}$ & 58.3(2) & A & \cite{vanUijen79} \\
Rb$_2$MnCl$_4$ (B $<$ 5.8 $\tesla$) & 0.15(1) & 1$\cdot$10$^{-1}-$1$\cdot$10$^{-3}$ & 54 & A & \cite{vandeKamp98,Tietze98} \\
K$_2$MnF$_4$ & 0.15(1) & 1$\cdot$10$^{-1}-$1$\cdot$10$^{-3}$ & 42.14 & A & \cite{Birgenau73} \\
K$_2$FeF$_4$ & 0.15(1) & $-$ & 63.0(3) & A & \cite{Thurlings89} \\
Rb$_2$MnF$_4$ & 0.16(2) & 1$\cdot$10$^{-1}-$3$\cdot$10$^{-3}$ & 38.4 & A & \cite{Birgenau70} \\
Pb$_2$Sr$_2$TbCu$_3$O$_8$ & 0.165(5) & $-$ & 5.30(2) & A & \cite{Wu96} \\
BaFeF$_4$ & 0.17 & 3$\cdot$10$^{-1}-$1$\cdot$10$^{-2}$ & & A & \cite{deJongh74} \\ %69.6(1)
Cr$_2$Si$_2$Te$_6$ & 0.17(1) & 6$\cdot$10$^{-1}-$3$\cdot$10$^{-2}$ & 32.1(1) & F & \cite{Carteaux95}\\
CsDy(MoO$_4$)$_2$ & 0.17(1) & $-$ & 1.36 & A & \cite{Khatsko04} \\
CoCl$_2\cdot$6H$_2$O & 0.18 & 4$\cdot$10$^{-1}-$4$\cdot$10$^{-2}$ & 2.29 & A & \cite{deJongh74} \\
MnC$_3$H$_7$PO$_3\cdot$H$_2$O$^{\dag}$ & 0.18(1) & 4$\cdot$10$^{-1}-$1$\cdot$10$^{-2}$ & $\sim15$ & F & \cite{Carling95} \\ %weak F
MnC$_4$H$_9$PO$_3\cdot$H$_2$O$^{\dag}$ & 0.18(1) & 4$\cdot$10$^{-1}-$2$\cdot10^{-2}$ & $\sim15$ & F & \cite{Carling95} \\ %weak F
KFeF$_4$ & 0.185(5) & 3$\cdot$10$^{-1}-$1$\cdot$10$^{-2}$ & 137.2(1) & A & \cite{Eibschutz72b,deJongh74} \\
Fe(NCS)$_2$(pyrazine)$_2$ & 0.19(2) & 2$\cdot$10$^{-1}-$3$\cdot$10$^{-2}$ & 6.8 & A & \cite{Bordallo04} \\
Rb$_2$FeF$_4$ & 0.2 & 3$\cdot$10$^{-1}-$2$\cdot$10$^{-3}$ & 56.3 & A & \cite{Birgenau70} \\
La$_2$CoO$_4$ & 0.20(2) & $-$ & 274.7(6) & A & \cite{Yamada89} \\
MnC$_2$H$_5$PO$_3\cdot$H$_2$O$^{\dag}$ & 0.21(2) & 6$\cdot$10$^{-1}-$9$\cdot$10$^{-2}$ & $\sim15$ & A & \cite{Carling95} \\
NH$_4$MnPO$_4\cdot$H$_2$O$^{\dag}$ & 0.21(3) & 8$\cdot$10$^{-1}-$2$\cdot$10$^{-2}$ & 17.5(1) & A & \cite{Carling95,Carling93} \\
K$_2$CuF$_4$ & 0.22 & 3$\cdot$10$^{-1}-$3$\cdot$10$^{-2}$ & 6.25 & F & \cite{Hirakawa73} \\
CuFo$_2\cdot$4D$_2$O$^{\ddag}$ & 0.22(2) & 5$\cdot$10$^{-1}-$5$\cdot$10$^{-2}$ & 16.72 & A & \cite{Koyama87} \\
CuFo$_2\cdot$CO(ND$_2$)$_2\cdot$2D$_2$O$^{\ddag}$ & 0.22(1) & 4$\cdot$10$^{-1}-$1$\cdot$10$^{-3}$ & 15.31 & A & \cite{Koyama87} \\
Tanol suberate$^{\S}$ & 0.22 & 7$\cdot$10$^{-1}-$2$\cdot$10$^{-2}$ & 0.7 & F & \cite{Blundell00} \\
Sr$_2$CuO$_2$Cl$_2$ & 0.22(1) & 2$\cdot$10$^{-1}-$1$\cdot$10$^{-2}$ & 265.5(5) & A & \cite{Greven95} \\
MnFo$_2\cdot$2H$_2$O & 0.22(1) & 4$\cdot$10$^{-1}-$4$\cdot$10$^{-2}$ & 3.6 & A & \cite{Matsuura85} \\
La$_2$NiO$_4$ & 0.22(2) & 8$\cdot$10$^{-2}-$2$\cdot$10$^{-3}$ & 327.5(5) & A & \cite{Nakajiama95} \\
BaNi$_2$(PO$_4$)$_2$ & 0.23 & 3$\cdot$10$^{-1}-$2$\cdot$10$^{-2}$ & 23.5(5) & A & \cite{Regnault90} \\
Cu(DCO$_2$)$_2\cdot$4D$_2$O & 0.23(1) & $t>6\cdot$10$^-2$ & 16.54(5) & A & \cite{Clarke92} \\
Rb$_2$CrCl$_4$ & 0.230(2) & 2$\cdot$10$^{-1}-$1$\cdot$10$^{-2}$ & 52.3 & F & \cite{Als-Nielsen93} \\
Gd$_2$CuO$_4$ & 0.23 & 7$\cdot$10$^{-1}-$3$\cdot$10$^{-3}$ & 6.4 & A & \cite{Chattopadhyay92}\\
(C$_6$H$_5$CH$_2$NH$_3$)$_2$CrBr$_4\,^{\P}$ & 0.23 & 7$\cdot$10$^{-1}-$1$\cdot$10$^{-1}$ & 52.0(1) & F & \cite{Bellitto87} \\
KMnPO$_4\cdot$H$_2$O$^{\dag}$ & 0.23(2) & $t>9\cdot$10$^{-2}$ & $\sim15$ & A & \cite{Carling95} \\
(CH$_3$NH$_3$)$_2$MnCl$_4$ & 0.23(2) & 1$\cdot$10$^{-2}-$1$\cdot$10$^{-3}$ & 44.75 & F & \cite{Paduan02} \\ %weak F
ErCl$_3$ & 0.23(2) & 4$\cdot$10$^{-1}-$1$\cdot$10$^{-2}$ & 0.350(5) & A & \cite{Kramer00} \\
(d$_6$-5CAP)$_2$CuBr$_4$ & 0.23(4) & 4$\cdot$10$^{-2}-$6$\cdot$10$^{-3}$ & 5.18(1) & A & \cite{Coomer07} \\
Li$_2$VOSiO$_4\,^{\S}$ & 0.235(9) & 4$\cdot$10$^{-1}-$2$\cdot$10$^{-2}$ & 2.85 & A & \cite{Melzi01} \\ %frustrated
Li$_2$VOGeO$_4\,^{\S}$ & 0.236 & $-$ & 1.95 & A & \cite{Carretta04} \\ %frustrated
La$_{0.04}$Sr$_{2.96}$Mn$_2$O$_7\,^{\S}$ & 0.24(2) & $-$ & 145.0(5) & A & \cite{Coldea02} \\ %frustrated
La$_{0.525}$Sr$_{1.475}$MnO$_4$ & 0.24(3) & $-$ & 110(1) & A & \cite{Larochelle05} \\ %frustrated
RbFeF$_4$ & 0.245(5) & 6$\cdot$10$^{-1}-$1$\cdot$10$^{-2}$ & 133(2) & A & \cite{deJongh74} \\
MnPS$_3$ & 0.25(1) & $t>3\cdot$10$^{-2}$ & 78.6 & A & \cite{Ronnow00,Wildes06} \\
Co$_5$(OH)$_8$(chdc)$\cdot$4H$_2$O & 0.25(3) & $-$ & 60.5 & F & \cite{Kurmoo03} \\ %ferrimagnet
YBa$_2$Cu$_3$O$_{6+x}$ & 0.26(1) & 5$\cdot$10$^{-2}-$5$\cdot$10$^{-3}$ & 410 & A & \cite{Montfrooij98} \\
Rb$_2$MnCl$_4$ (B $>$ 5.8 $\tesla$) & 0.26(1) & 1$\cdot$10$^{-1}-$2$\cdot$10$^{-3}$ & 54 & A & \cite{Tietze98,vandeKamp98} \\
Rb$_2$CrCl$_3$Br & 0.26(1) & 9$\cdot$10$^{-1}-$1$\cdot$10$^{-2}$ & 55 & F & \cite{Bramwell89,Bramwell86} \\
Rb$_2$CrCl$_2$Br$_2$ & 0.26(1) & 9$\cdot$10$^{-1}-$3$\cdot$10$^{-2}$ & 57 & F & \cite{Bramwell89,Bramwell86} \\
KMnF$_4$ & 0.26(1) & 3$\cdot$10$^{-1}-$3$\cdot$10$^{-2}$ & 5.2(1) & A & \cite{Moron93} \\
RbMnF$_4$\ & 0.26(1) & 3$\cdot$10$^{-1}-$3$\cdot$10$^{-2}$ & 3.7(1) & A & \cite{Moron93} \\
\hline
\hline
\multicolumn{6}{l}{{\rule[-0mm]{0mm}{5mm}}$^{\ast}$ Studied by M\"ossbauer spectroscopy.} \\
\multicolumn{6}{l}{$^{\dag}$ Studied by bulk magnetometry.} \\
\multicolumn{6}{l}{$^{\ddag}$ Studied by proton Nuclear Magnetic Resonance (NMR).} \\
\multicolumn{6}{l}{$^{\S}$ Studied by muon Spin Rotation ($\mu$SR).} \\
\multicolumn{6}{l}{$^{\P}$ Studied by ac susceptibility.}
\end{tabular}
\label{layered_table}
\normalsize
\end{table}
{\footnotesize
\begin{landscape}

\begin{longtable}{lcccccccc}
\caption{Summary of transition temperatures $T_{\mathrm{c}}$ and magnetisation critical exponents $\beta$ for epitaxial magnetic films grown on a range of substrates. The thickness $d_{\mathrm{min}}$ denotes the thickness at which these quantities were measured and $t$ range indicates the range of reduced temperature over which $\beta$ was measured. The magnetic anisotropy is indicated by the direction of the easy axis, and can either be perpendicular ($\perp$) or parallel ($\parallel$) to the film plane.} \label{thinfilm_table} \\

\hline
\hline 
\multicolumn{1}{l}{System} & \multicolumn{1}{c}{Structure} & \multicolumn{1}{c}{$d_{\mathrm{min}}$ (ML)} & \multicolumn{1}{c}{$\beta$} & \multicolumn{1}{c}{$t$ range} & \multicolumn{1}{c}{$T_{\mathrm{c}}$ ($\kelvin$)} & \multicolumn{1}{c}{Anisotropy} & \multicolumn{1}{c}{Method\footnotemark[1]} & \multicolumn{1}{c}{Reference} \\ \hline 
\endfirsthead

\multicolumn{9}{c}%
{{\slshape \tablename\ \thetable{} -- continued from previous page}} \\
\hline 
\hline
\multicolumn{1}{l}{System} & \multicolumn{1}{c}{Structure} & \multicolumn{1}{c}{$d_{\mathrm{min}}$ (ML)} & \multicolumn{1}{c}{$\beta$} & \multicolumn{1}{c}{$t$ range} & \multicolumn{1}{c}{$T_{\mathrm{c}}$ ($\kelvin$)} & \multicolumn{1}{c}{Anisotropy} & \multicolumn{1}{c}{Method\footnotemark[1]} & \multicolumn{1}{c}{Reference} \\ \hline 
\endhead

\hline \multicolumn{9}{r}{{\slshape Continued on next page}} \\ \hline
\endfoot

\hline \hline
\endlastfoot

Fe on & & & & & & & & \\
Pd(100) & bct, $1\times1$ & 2.0 & 0.125(10) & $t<3\cdot10^{-2}$ & 613.1 & $\perp$ & ECS & \cite{Rau93} \\
%        & 0.7 & 0.15(1) & 3$\cdot$10$^{-1}-$1$\cdot$10$^{-2}$ & $>100$ & $\perp$ & MOKE & \cite{Liu1990} \\
        & & 1.2 & 0.127(4) & 1$\cdot$10$^{-1}-$3$\cdot$10$^{-3}$ & $<100$ & $\perp$ & MOKE & \cite{Liu90} \\
Ag(100) & bcc, $1\times1$ & 2.5-2.7\footnotemark[2] & 0.124(2) & 1$\cdot$10$^{-1}-$1$\cdot$10$^{-3}$ & 324 & $\perp$ & MOKE & \cite{Qiu93,Qiu94} \\
W(110)  & bcc, $1\times1$ & 0.8 & 0.124(1) & 1$\cdot$10$^{-1}-$4$\cdot$10$^{-3}$ & 221.1(1) & $\parallel$ [$1\overline{1}0$] & SPLEED & \cite{Elmers94,Elmers95} \\
        & & 1.0 & 0.134(3) & 1$\cdot$10$^{-1}-$5$\cdot$10$^{-2}$ & 224 & $\parallel$ [$1\overline{1}0$] & SPLEED & \cite{Elmers96_Ising} \\
        & & 1.7 & 0.13(2) & $-$ & 317 & $\parallel$ [$1\overline{1}0$] & MOKE & \cite{Back95} \\
Ag(111) & bcc, $1\times1$ & 1.8 & 0.139(6) & 1$\cdot$10$^{-1}-$1$\cdot$10$^{-3}$ & $\sim450$ & $\parallel$ & MOKE & \cite{Qiu91} \\
        & & 2.0 & 0.130(3) & 1$\cdot$10$^{-1}-$1$\cdot$10$^{-3}$ & $\sim450$ & $\parallel$ & MOKE & \cite{Qiu91} \\
Cu(100) & fct, $4\times1$ & $\sim2.5$\footnotemark[2] & 0.17(3) & 1$\cdot$10$^{-1}-$1$\cdot$10$^{-2}$ & 370 & $\perp$\footnotemark[3] & MOKE & \cite{Li94,Thomassen92,Pappas90} \\
W(110)\footnotemark[4] & bcc, $1\times1$ & 0.82 & 0.18(1) & 3$\cdot$10$^{-1}-$1$\cdot$10$^{-2}$ & 282(3) & $\parallel$ [$1\overline{1}0$] & TOM, CEMS & \cite{Gradmann89,Przybylski87,Elmers89} \\
Cu$_{84}$Al$_{16}$(100) & fcc, $1\times1$ & 4.0 & 0.212(5) & 3$\cdot$10$^{-1}-$1$\cdot$10$^{-2}$ & 288(2) & $\parallel$ & LMDAD & \cite{Macedo98} \\
W(100)  & bcc, $1\times1$ & 1.6 & 0.217(2) & 1$\cdot$10$^{-1}-$1$\cdot$10$^{-2}$ & 207.8(1) & $\parallel$ [$001$] & CEMS, SPLEED & \cite{Elmers96_XY,Elmers94b} \\
Au(100) & bcc, $1\times1$ & 1.0 & 0.22(1) & 1$\cdot$10$^{-1}-$1$\cdot$10$^{-3}$ & 300 & $\parallel$ [$001$] & SPLEED & \cite{Durr89} \\
        & & 2.0 & 0.25(1) & 2$\cdot$10$^{-1}-$1$\cdot$10$^{-4}$ & 290 & $\parallel$ [$001$] & ECS & \cite{Rau89} \\
W(100)\footnotemark[4] & bcc, $1\times1$ & 1.5 & 0.22(2) & $-$ & 282(1) & $\parallel$ [$001$] & CEMS & \cite{Elmers95} \\
V(001) & bcc & 3 & 0.23(1) & 2$\cdot$10$^{-1}-$2$\cdot$10$^{-2}$ & $\sim190$ & $\parallel$ & MOKE & \cite{Parnaste05b} \\
Pd\footnotemark[5] & $-$ & 0.2-0.4\footnotemark[2] & 0.23(1) & 2$\cdot$10$^{-1}-$2$\cdot$10$^{-2}$ & $>50$ & $\parallel$ & MOKE & \cite{Parnaste07} \\
GaAs(100) & bcc, $2\times6$ & 3.4 & 0.26(2) & 1$\cdot$10$^{-1}-$1$\cdot$10$^{-3}$ & 254.8(2) & $\parallel$ & MOKE & \cite{Bensch01} \\
\hline
Co on & & & & & & & \\
Cu(111) & fcc, $1\times1$ & 1.0 & 0.125 & $-$ & 433 & $\perp$ & TOM & \cite{Kohlepp92} \\
        & & 1.5 & 0.15(8) & $-$ & 460 & $\perp$ & MOKE & \cite{Huang94} \\
Cu(100) & fcc & 2.0 & 0.24 & $-$ & $\sim240$ & $\parallel$ & MOKE & \cite{Kuo02,Gruyters05} \\
\hline
\pagebreak Ni on & & & & & & & \\
W(110)  & fcc, $7\times1$ &  2.0 & 0.13(6) & 1$\cdot$10$^{-1}-$1$\cdot$10$^{-3}$ & 325 & $\parallel$ [$001$] & FMR & \cite{Li92} \\
Cu(111) & fcc, $1\times1$ & 2.0-3.0\footnotemark[2] & 0.24(7) & 3$\cdot$10$^{-1}-$6$\cdot$10$^{-3}$ & 380 & $\parallel$ & MOKE & \cite{Ballantine90} \\
Cu(100) & fcc, $1\times1$ & 4.1 & 0.23(5) & 3$\cdot$10$^{-1}-$1$\cdot$10$^{-2}$ & 284 & $\parallel$ & MOKE & \cite{Huang93,Huang94} \\
\hline
V on & & & & & & & \\
Ag(100) & bcc, $1\times1$ & 5.0 & 0.128(10) & 3$\cdot$10$^{-1}-$2$\cdot$10$^{-4}$ & 475.1 & $\parallel$ [$001$] & ECS & \cite{Rau88} \\
\hline
Gd on & & & & & & & \\
Y(0001) & hcp & 1.0 & 0.23(5) & 1$\cdot$10$^{-1}-$8$\cdot$10$^{-3}$ & 156 & $\parallel$ & MOKE & \cite{Gajdzik98} \\
\hline
Mn$_5$Ge$_3$ on & & & & & & & \\
Ge(111) & & 1.0 & 0.244 & 2$\cdot$10$^{-1}-$4$\cdot$10$^{-3}$ & 296 & $\parallel$ & SQUID & \cite{Zeng03} \\
\hline
CoAl(100) & bcc, $1\times1$ & & 0.22(2) & 2$\cdot$10$^{-1}-$7$\cdot$10$^{-3}$ & $\sim90$ & $\parallel$ & MOKE & \cite{Rose07}

\end{longtable}
\footnotetext[1]{Experimental properties were measured by the following techniques: Electron Capture Spectroscopy (ECS), Magneto Optical Kerr Effect (MOKE), Spin Polarised Low Energy Electron Diffraction (SPLEED), Torsion Oscillation Magnetometry (TOM), Convertion Electron M\"ossbauer Spectroscopy (CEMS), Linear Magnetic Dichroism in the Angular Distribution of photoelectron intensity (LMDAD) and Superconducting QUantum Interference Device (SQUID).}
\footnotetext[2]{Exponent determined by averaging over values of a range of films of different thickness.}
\footnotetext[3]{Reversible spin reorientation transition from $\parallel$ to $\perp$ with increasing $T$.}
\footnotetext[4]{Coated with Ag.}
\footnotetext[5]{Pd layers $\delta$-doped with Fe.}

\end{landscape}
}

\begin{table}
\footnotesize
\caption{Chemisorbed and physisorbed systems displaying two-dimensional phase transitions.}
\vspace{1mm}
\begin{tabular}{lcccc}
\hline 
\hline
System & $\beta$ & Model Ascribed & Method & Reference \\
\hline
W(011)$p(2\times1)$-H & 0.13(4) & 2d Ising$^{\ast}$ & LEED$^{\dag}$ & \cite{Lyuksyutov81} \\
W(011)$p(2\times2)$-H & 0.25(7) & 2dXY$^{\ast}$ & LEED & \cite{Lyuksyutov81} \\
$p(1\times2)\leftrightarrow(1\times1)$-Au(110) & 0.13(2) & 2d Ising & LEED & \cite{Capunzano85} \\
W(112)$p(2\times1)$-O & 0.13(1) & 2d Ising & LEED & \cite{Wang85} \\
$p(2\times2)$-O disordering on Ru(001) & 0.13(2) & 3-state Potts & LEED & \cite{Pfnur90} \\
$p(2\times1)$-O disordering on Ru(001) & 0.085(15) & 4-state Potts & LEED & \cite{Pfnur89} \\
$p(2\times2)$-O disordering on Mo(110) & 0.19(2) & XY$h_4^{\ast}$ & LEED & \cite{Grzelakowski90} \\
$p(2\times1)$-O disordering on W(110) & 0.19(5) & XY$h_4^{\ast}$ & LEED & \cite{Baek93} \\
Ru(001)$p(2\times2)$-S & 0.11(2) & 4-state Potts & LEED & \cite{Sokolowski94} \\
Ru(001)$(\sqrt{3}\times\sqrt{3})R30^{\circ}$-S & 0.14(3) & 3-state Potts & LEED & \cite{Sokolowski94} \\
$(3\times3)$-Sn disordering on Ge(111) & 0.11(1) & 3-state Potts & HAS$^{\ddag}$, XRD$^{\S}$ & \cite{Floreano01} \\
$(3\times1)$ reconstruction on Si(113) & 0.11(4) & 3-state Potts & LEED & \cite{Yang90} \\
Xe melting on graphite & 0.23(4) & 2dXY & XRD & \cite{Nuttall95} \\
\hline
\hline
\multicolumn{4}{l}{{\rule[-0mm]{0mm}{5mm}}$^{\ast}$ Model not ascribed by original authors.} \\
\multicolumn{4}{l}{$^{\dag}$ LEED: Low Energy Electron Diffraction.} \\
\multicolumn{4}{l}{$^{\ddag}$ HAS: Helium diffraction.} \\
\multicolumn{4}{l}{$^{\S}$ XRD: X-ray diffraction.} \\
\end{tabular}
\label{orderdisorder}
\normalsize
\end{table}

%\section{Extra points}

% Caption and footnotes in Table A2 need to be formatted correctly. This is a problem to do with the {\tt longtable} package. Surely the JPCM editor can offer advice on this?

\clearpage

\bibliographystyle{unsrt}
\bibliography{references}

\end{document}